\documentclass[11pt,twoside]{article}

\usepackage{asp2006}
\usepackage{epsf}
\usepackage{psfig}
\usepackage{lscape}
\usepackage{graphicx}

\begin{document}
\title{Impact of low-frequency p modes on the inversions of the internal rotation of the Sun}   %%% Fill in title
\author{D. Salabert$^1$, A. Eff-Darwich$^{1,2}$, R. Howe$^3$, and S.~G. Korzennik$^4$}   %%% Fill in author names

\affil{$^{1}$Instituto de Astrof\'isica de Canarias, C/ V\'ia L\'{a}ctea s/n, 38205, La Laguna, Tenerife, Spain}   
\affil{$^{2}$Departmento Edafolog\'ia y Geolog\'ia, Universidad de La Laguna, La Laguna, Tenerife, 38205, Spain}
\affil{$^{3}$National Solar Observatory, 950 North Cherry Avenue, Tucson, AZ~85729, USA}
\affil{$^{4}$Harvard-Smithsonian Center for Astrophysics, 60 Garden Street, Cambridge, MA~02138, USA}

\begin{abstract} %%% Abstract to run on from here.
We used the $m$-averaged spectrum technique (``collapsogram'') to extract the low-frequency solar p-mode 
parameters of low- and intermediate-angular degrees ($l \leq 35$) in long time series of GONG and MDI observations. 
Rotational splittings and central frequencies have been measured down to $\approx$~850~$\mu$Hz, including predicted 
modes which have not been measured previously. Both GONG and MDI frequency splitting data sets were numerically inverted 
to extract the internal solar rotation rate. The impact of the very low-frequency observables and the differences 
between GONG and MDI data sets on the inversion results are also analyzed.
\end{abstract}

\section{Measurements of low-frequency p modes of low- and medium-angular degrees ($1 \leq l \leq 35$)}
\begin{figure}
\centering
\includegraphics[angle=90,width=0.9\textwidth]{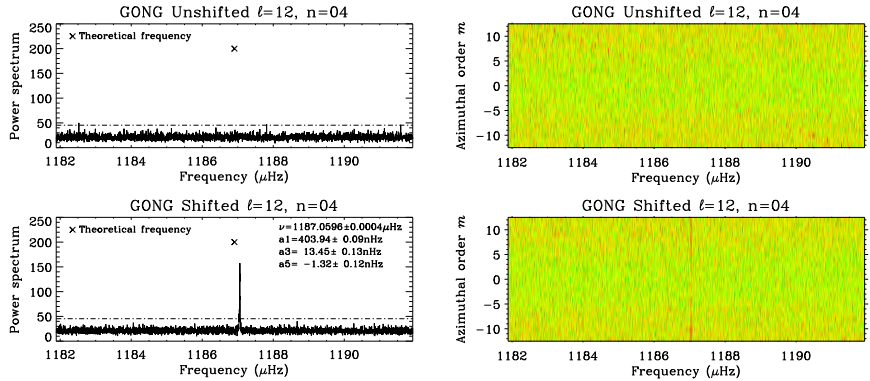}
\caption{Collaspograms for $l=12, n=4$, and their $m-\nu$ diagrams before
and after shifting in frequency by the indicated estimates of the $a$-coefficients.}
\label{fig:colps}
\end{figure}

\begin{figure}
\centering
\includegraphics[width=0.8\textwidth]{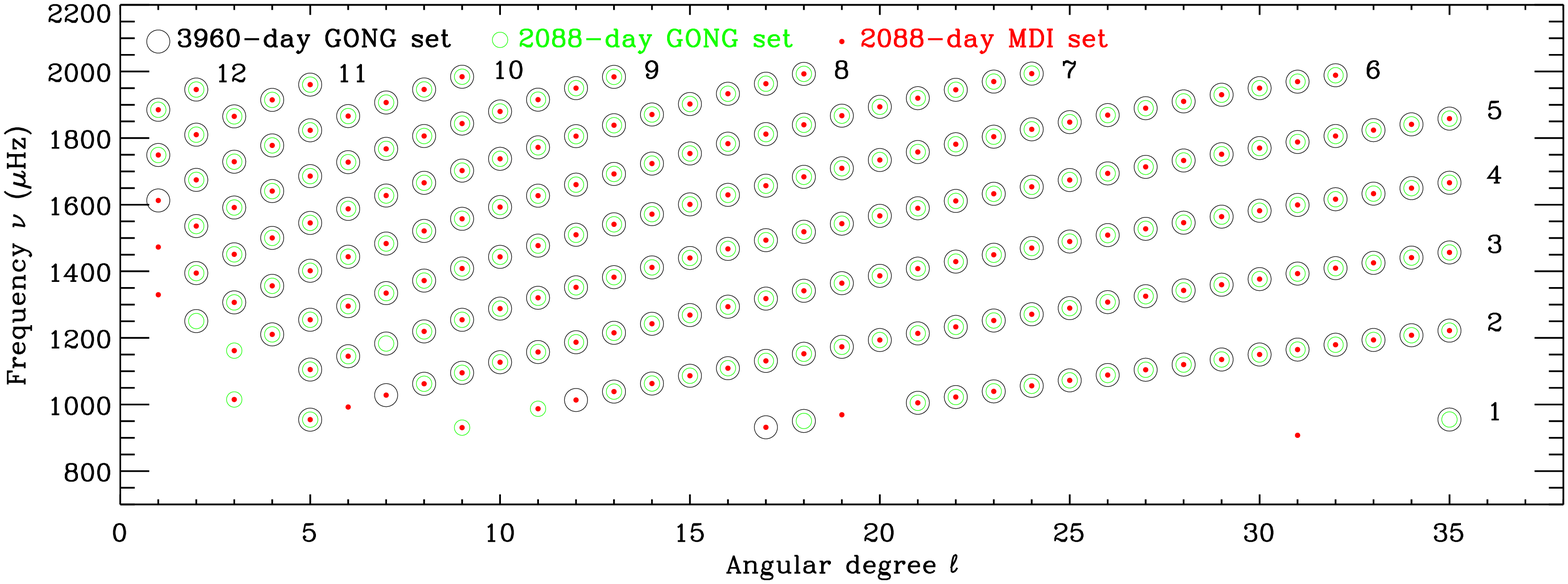}
\caption{$l-\nu$ diagram of the low-frequency modes ($l\leq35$) observed with the collapsograms
in 3960 days of GONG observations (black) and 2088 coeval days of GONG (green) and MDI (red) 
observations. The ridges of same radial order
are also indicated from $n=1$ to $n=12$.}
\label{fig:lnu}
\end{figure}

In the search of low-frequency acoustic modes, the usual peak-fitting methods, using individual-$m$ spectra, 
are naturally limited by the decreasing signal-to-noise ratio (SNR). Instead, several pattern-recognition 
techniques have been developed in an effort to reveal the presence of the modes in the low-frequency range.
We use here an adaptation of the rotation-corrected $m$-averaged spectrum technique which finds the best set of 
the $a$-coefficient splittings yielding the narrowest profile in the average spectrum (Salabert, Leibacher, \& Appourchaux~2007). 
Before averaging, each $m$-spectrum at a given ($n, l$) is shifted by a frequency that compensates for the effect 
of differential rotation and non-spherical effects on the frequencies. A high SNR can result from combining individual 
low-SNR individual-$m$ spectra, none of which would yield a strong enough peak to measure (Fig.~\ref{fig:colps}). 
This method, called collapsogram, 
has been applied to 3960 days of GONG data and to 2088-day coeval observations of GONG and MDI for modes with $1 \leq l  \leq 35$. 
Acoustic modes down to $\approx 850 \mu$Hz have been observed (Fig.~\ref{fig:lnu}) and their parameters estimated by fitting the 
rotation-corrected $m$-averaged spectrum with an asymmetric Lorentzian profile.

\section{Rotational inversions of the GONG and MDI low-frequency p~modes measured using the collapsograms}
\begin{figure}
\centering
\plottwo{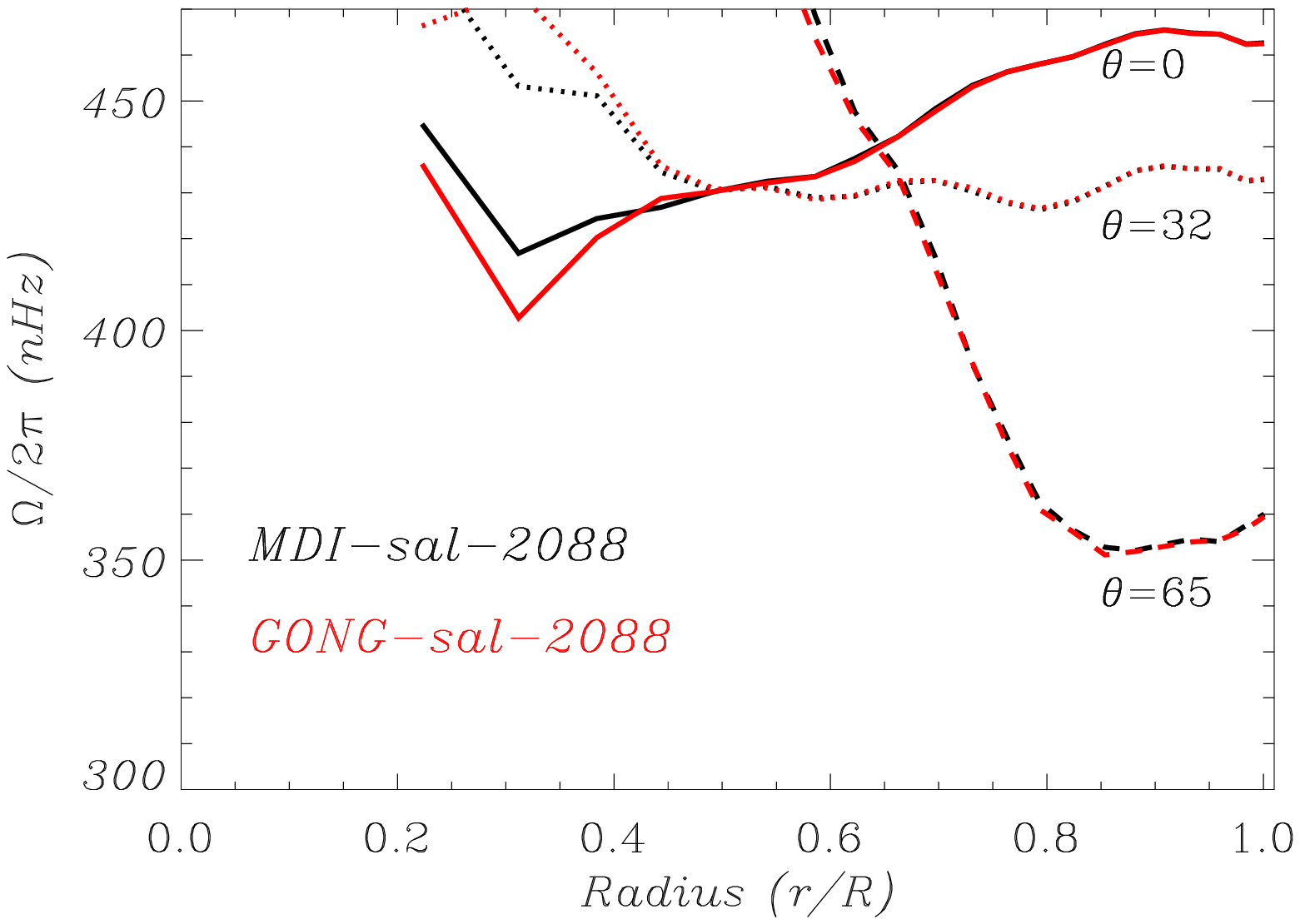}{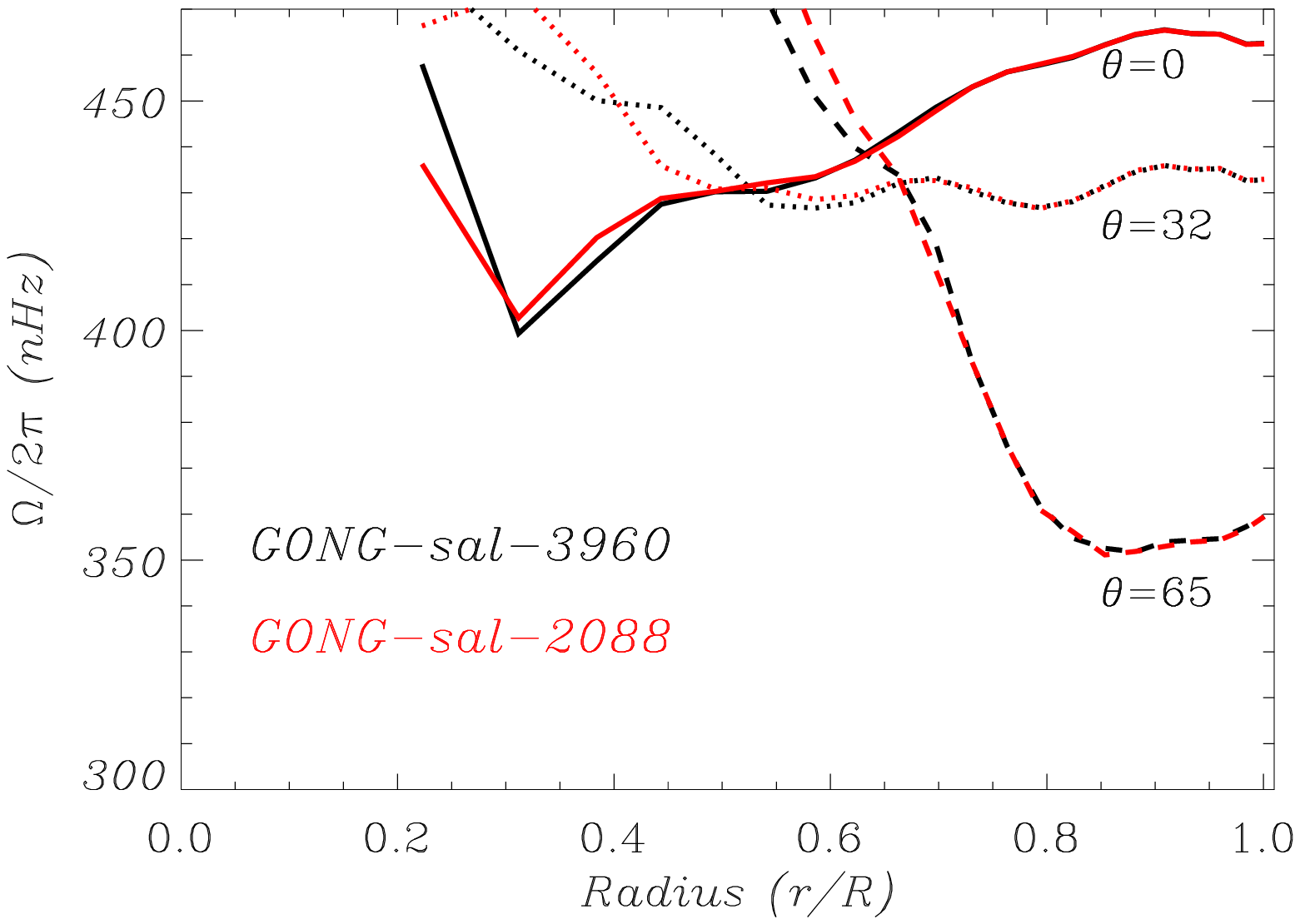}
\caption{Rotational inversions of the low-frequency p modes with $l \leq 35$ measured using the collapsograms for
2088-day GONG and MDI data sets (left panel), and 2088-day and 3960-day GONG data sets (right panel).}
\label{fig:inv0}
\end{figure}

\begin{figure}
\centering
\includegraphics[width=0.6\textwidth]{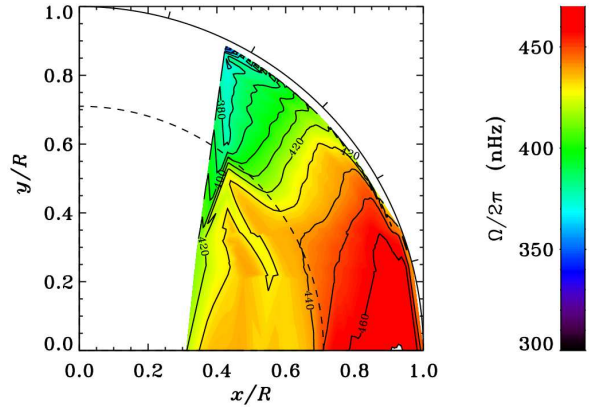}
\caption{Inversion of the 3960-day GONG data set combined with the mean of 35~$\times$~108-day GONG PEAKFIND data.}
\label{fig:rachel}
\end{figure}

Rotational inversions were computed using the $a$-coefficients of the low-frequency p modes 
of low- and intermediate-angular degrees measured with the collapsograms in long times series 
of GONG and MDI observations (Fig.~\ref{fig:inv0}). The inversions were carried out using an iterative 
method implemented to avoid the need to invert matrices (recalling the ill-posed nature of 
the inversion problem). For details about the inversion methodology, see Eff-Darwich \& Korzennik (these proceedings). 
Only modes with $l \leq 35$ and $\nu \leq$~2100~$\mu$Hz were used. 
The inversions of both MDI and GONG data sets give consistent results and no significant differences 
are found when inverting 2088 days or 3960 days of data. Since observational 
uncertainties are reduced by a factor proportional to the squared root of the number of days, it will be necessary 
to have extremely long series to find significant improvements in the rotational distribution. 
Figure~\ref{fig:rachel} shows the rotational inversion of the low-frequency modes measured with the collapsograms
combined with the mean of 35~$\times$~108-day GONG PEAKFIND data (using both the higher-degree and higher-frequency 
modes from PEAKFIND).

\section{Comparisons between the collapsograms and other measurements}
\begin{figure}
\centering
\plottwo{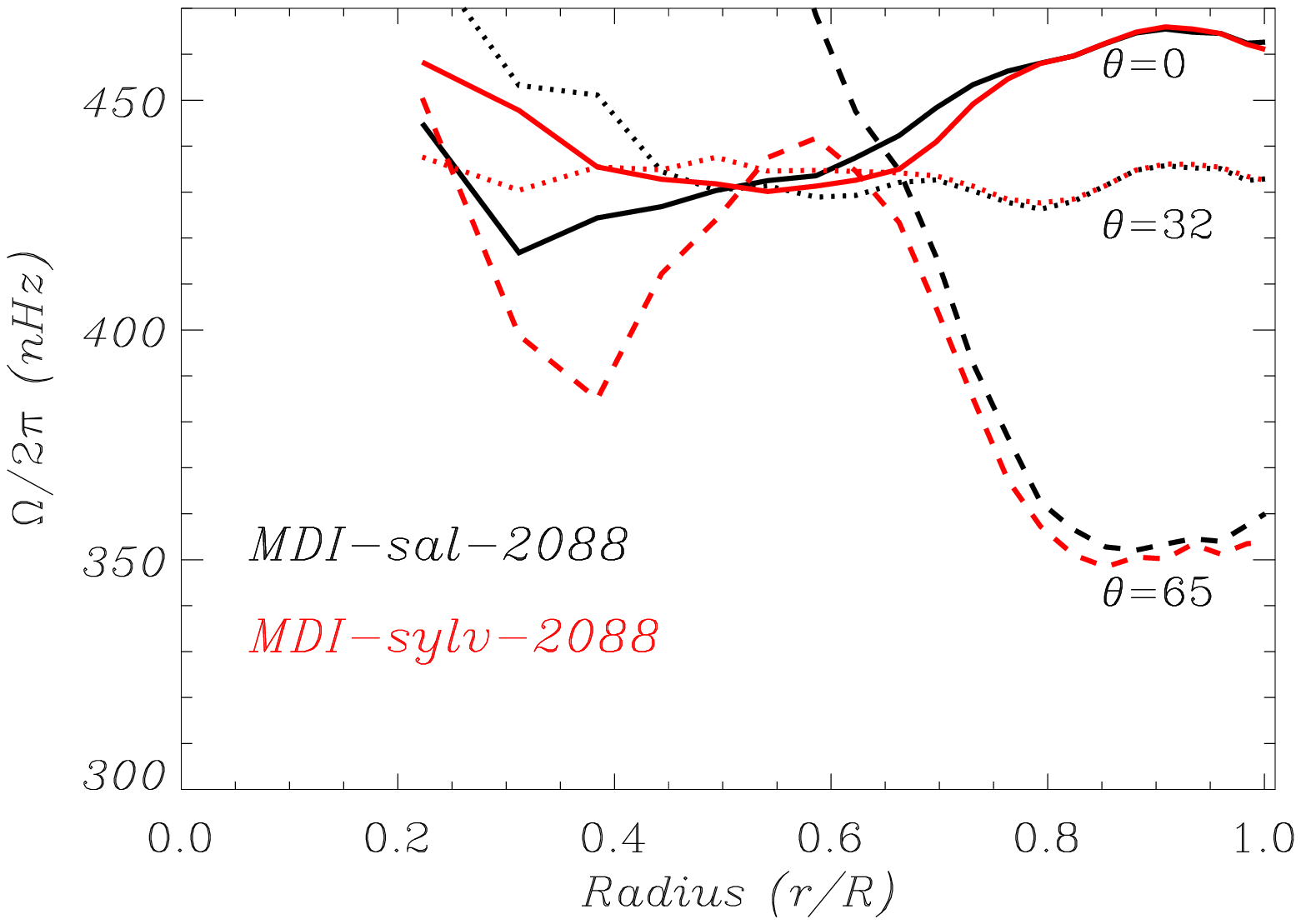}{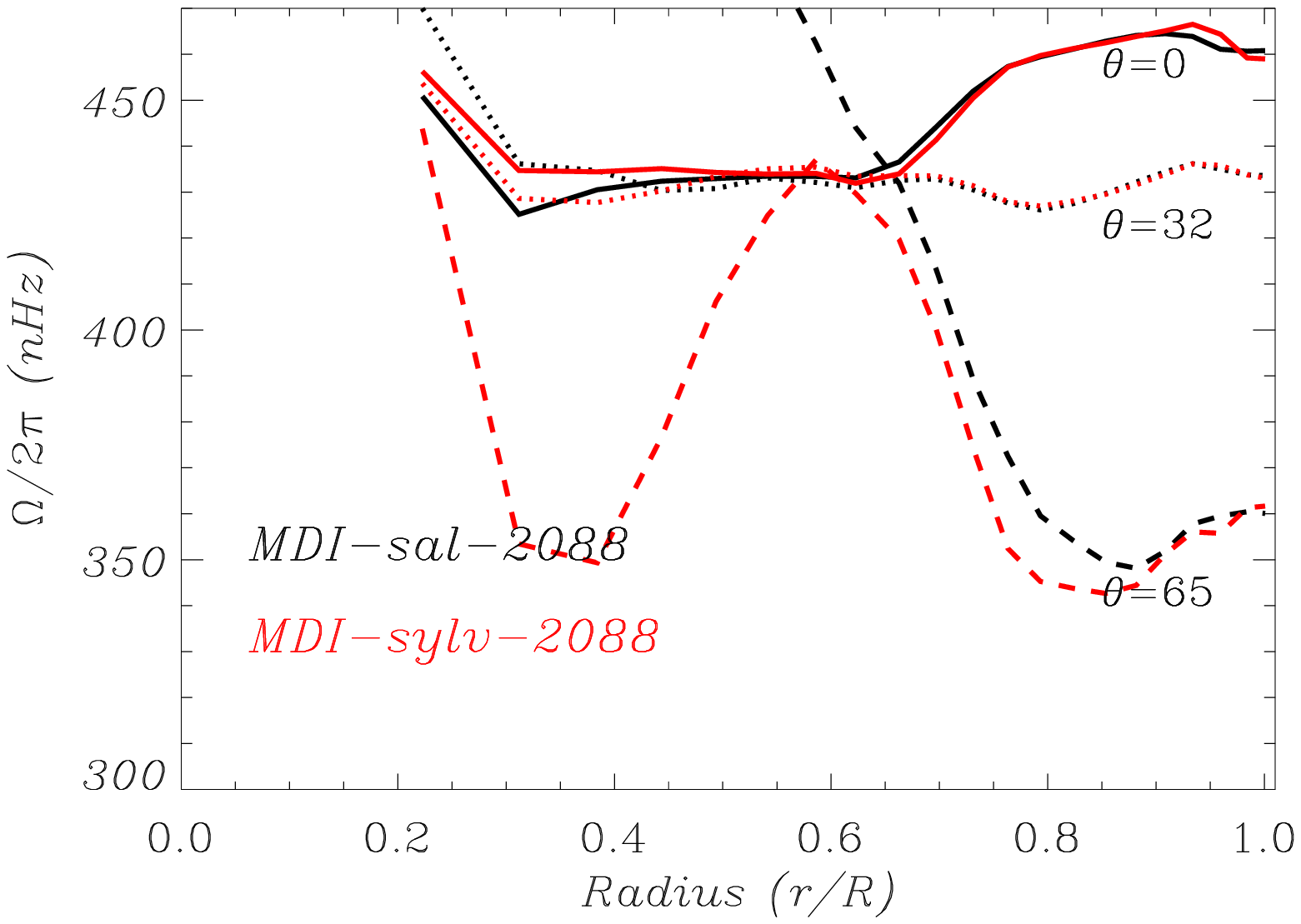}
\caption{Rotational inversions of the MDI 2088-day data sets of p modes with 
$l \leq 35$ and $\nu \leq$~ 2100~$\mu$Hz obtained using the collapsograms and by Korzennik (2005) (left panel). 
Same as left panel but complemented with the higher-degree and higher-frequency p modes from Korzennik (2005) (right panel).
}
\label{fig:inv1}
\end{figure}

Figure~\ref{fig:inv1} shows the comparisons between the rotational inversions using the $a$-coefficients 
obtained with the collapsograms and those measured by Korzennik (2005) fitting the individual-$m$ spectra. 
In both methods, the same coeval 2088 days of observations have been used.
The effect of the differences in the low-frequency ($\nu \leq$~2100~$\mu$Hz) and low-degree ($l \leq 35$) 
splitting data sets between Korzennik (2005) and the ones obtained using the collapsograms
 are translated to the internal rotational profiles. Special care should be taken regarding the
 calculation of the observational errors when only Clebsch-Gordan coefficients are available 
(Eff-Darwich \& Korzennik, these proceedings). Small differences in the data sets result in the observed large 
discrepancies in the radiative zone at high latitude, due to a  lack of sensitivity in these regions
(right panel of Fig.~\ref{fig:inv1}).

\section{Conclusions}
We performed rotational inversions by using the $a$-coefficients of the low-frequency solar p modes of 
low- and intermediate-angular degrees ($1 \leq l \leq 35$) measured in long time series of GONG and MDI observations with the 
$m$-averaged spectrum technique (also called collapsogram, Salabert et al. 2007).
We showed that the inversions of both GONG and MDI data sets give consistent results and that no significant differences are found
when inverting 2088 days or 3960 days of data. We also compared with the rotational inversions obtained by using the $a$-coefficients
measured by Korzennik (2005) fitting the individual-$m$ spectra and stressed out the importance on how the observational errors are calculated.

\acknowledgements %%% Text of acknowledgements runs on after this command.
This work utilizes data obtained by the Global Oscillation Network Group (GONG) 
Program, managed by the National Solar Observatory, which is operated by AURA, Inc. under 
a cooperative agreement with the National Science Foundation. The data were acquired by 
instruments operated by the Big Bear Solar Observatory, High Altitude Observatory, 
Learmonth Solar Observatory, Udaipur Solar Observatory, Instituto de Astrof\'isica de Canarias, 
and Cerro Tololo Interamerican Observatory. The MDI instrument on-board SOHO is a 
cooperative effort to whom we are indebted. SOHO is a project of international collaboration 
between ESA and NASA. D.~S. acknowledges the support of the Spanish grant PNAyA2007-62650.

\end{document}